\documentclass[letter]{aa} 

\usepackage{graphicx}
\usepackage{natbib}
\usepackage{epsfig}
\usepackage{txfonts}
\begin{document}

   \title{The fraction of BL Lac objects in groups of galaxies}


\author{H. Muriel \inst{1} }

 \institute{Observatorio Astron\'omico de C\'ordoba (UNC) and Instituto de Astronom\'ia Te\'orica y Experimetal 
(CONICET-UNC). C\'ordoba, Argemtina.\\
              \email{hernan@oac.unc.edu.ar}}

   \date{Received ..., ; accepted ..., }


  \abstract
   {BL Lac objects are a rare class of active galactic nuclei that typically show featureless optical spectra that make it difficult to
estimate the redshift. A novel method  for estimating the redshift of BL Lac objects has recently been proposed which assumes that 
these objects typically reside in groups of galaxies.   }
{  The aim of this work is to estimate the fraction of BL Lac objects that reside in groups of galaxies. 
   }
   {We use a sample of groups (M12 catalogue) selected by applying a {friends-of-friends} algorithm in the SDSS DR12. 
Galaxies in the M12 sample were cross-correlated with the sample of blazars in the BZCAT.
We found that 121 galaxies in the M12 catalogue are blazars in the BZCAT sample, all but one are BL Lac objects, and 
a large fraction are classified as {BL Lac-galaxy dominated}.
   }
   {Analysing the fraction of galaxies in groups as a function of redshift we have 
estimated a correction factor that takes into account the typical incompleteness of the catalogues based on the  
{friends-of-friends} algorithm. Once this factor was
applied to the sample of BL Lac objects with a counterpart in the M12 catalogue, we found that the percentage of BL Lac objects in 
groups is $\gtrsim 67 \pm 8\%$. 
   }
   {The high rate of BL Lac objects in groups found in this work strongly supports a recent method that has    successfully    estimated the redshift of BL Lac objects with featureless spectra.
 }


   \maketitle
%

\section{Introduction}

A blazar is a type of active galactic nucleus (AGN) typically associated with elliptical galaxies \citep{Falomo:1996}.
Blazars are 
one of the most energetic phenomena in the universe.  BL Lac objects are a very specific subtype of blazar
that show a rapid and large-amplitude flux variability and have radio-loud AGNs. According 
to the unified scheme for AGNs (\citet{Antonucci:1993}, \citet{Urry:1995}), the strong emission of BL Lac objects is interpreted 
as the result of a relativistic jet   aligned with the line of sight of the observer. BL Lac objects are of great interest 
to the high-energy astrophysics community since they have a great potential to constrain the models for the extragalactic 
background light (EBL).

A significant fraction of BL Lac objects have featureless 
spectra, which makes it difficult or impossible to estimate the redshift of the host galaxy. Approximately 50\% of
the known BL Lac objects do not have redshift estimates.
\citet{Muriel:2015} have proposed a novel method for  estimating the redshift of BL Lac objects. Under the assumption that these AGNs
are hosted by early-type galaxies and that these in turn tend to live in systems of galaxies, they have proposed to study the
environment of BL Lac objects in order to find their host group/cluster of galaxies. The method was first applied to 
PKS 0447-439 where these authors found a group of galaxies with at least seven members at z=0.343 and estimated a
probability of $\geq 97\%$ that the host galaxy of PKS 0447-439 was a member of this group. The same method has been 
applied by \citet{Rovero:2016} to  PKS 1424+240, who have found a group of eight or more members at a redshift of 0.0601 and a probability of 
$\geq 98\%$ that this group hosts this BL Lac. The proposed redshift of $\sim 0.6$ for PKS 1424+240 found by \citet{Rovero:2016} 
is consistent with other methods used to constrain the redshift of this BL Lac. Modelling the change produced by the EBL from 
high to very high energy in the spectral index of this BL Lac object, \citet{Acciari:2010} have found for 
PKS 1424+240 an upper limit of $z < 0.66$ (see also \citet{Aleksic:2014}, \citet{Yan:2015a} and \citet{Yan:2015b}). 
By analysing the Ly$\beta$ and Ly$\gamma$ absorption in the far UV, \citet{Furniss:2013} have found  a lower 
limit of $z > 0.6035$ for the same object. The high consistency between these upper and lower limits with the redshift found by \citet{Rovero:2016} 
strongly supports the method proposed by \citet{Muriel:2015}. 

Even in the case that only one group of galaxies is found in the line of sight of a given BL Lac object, the technique applied by 
\citet{Muriel:2015} and \citet{Rovero:2016} 
requires an estimate of the probability that a BL Lac object is in a group of galaxies. These authors estimated that at least 2/3 of
the BL Lac objects are in systems of galaxies, but currently  there is no accurate estimate of this fraction.  During the nineties, 
several studies addressed the environment of BL Lac objects. \citet{Wurtz:1993} studied the environment around five BL Lac objects 
and found evidence of rich clusters or groups of galaxies around four of them. Using both imaging and spectroscopy, 
\citet{Pesce:1994} found an environment around four BL Lac objects similar to that of Abell clusters of richness class 0. Similar results 
were found by \citet{Smith:1995}. Based on images from the Hubble Space Telescope, \citet{Pesce:2002} found, for a deep sample  
of BL Lac objects, that the environment of this type of AGN tends to show a galaxy enhancement over the background. Nevertheless, 
for some of the objects in their sample, they also found possible evidence that some BL Lac objects can be truly isolated. A group/cluster 
environment has been reported around individual BL Lac objects (\citet{Lietzen:2008}).
Owing to both the small size of the samples and the limitations of the techniques used to characterize the environment of
BL Lac objects, it is quite difficult to make a reliable prediction of the fraction of BL Lac objects that inhabit 
groups/clusters of galaxies.

Using a large sample of groups of galaxies, in this paper we cross-correlate single galaxies and group members 
with a large compilation of blazars. The aim of this work is to obtain a precise estimate of the fraction of
BL Lac objects that inhabit systems of galaxies. This paper is organized as follows: In section 2 we describe the sample of groups and
blazars used to cross-correlate; the results of this analysis are presented in section 3, including the procedure applied to
correct for incompleteness;  conclusions are presented in section 4. Distances have been computed assuming a flat cosmological 
model with parameters $H_0 = 100 h km s^{-1} Mpc^{-1}$,
$\Omega_0 = 0.3$, and $\Omega_{\Lambda} = 0.7$.

\section{Sample}
\subsection{Groups of galaxies}
We use an updated version of the catalogue of groups of galaxies obtained by \citet{Merchan:2005} using the SDSS. 
The original catalogue was based on the Third Data Release (DR3, \citet{Abazajian:2005}), while the new version compiles 
galaxies and groups in the DR12 (\citet{Adam:2015}). Despite the difference in the source catalogue,
 the new sample was compiled (by Merch\'an, private communication, hereafter M12) by applying the same procedure 
described in \citet{Merchan:2005}. Group identification is based on the {friends-of-friends} (FOF) algorithm developed 
by \citet{Huchra:1982}, which takes into account the density variation produced by the apparent magnitude limit 
of redshift surveys. The method
sets two linking lengths, one in the line of sight ($V_0 R$) and one perpendicular to the plane of the sky ($D_0 R$), 
where $R$ is the scaling factor. The value of $D_0$ is chosen to obtain a certain overdensity $\delta \rho/\rho$. 
M12 includes
two samples of groups of galaxies corresponding to overdensities of 80 and 200, respectively. In order to minimize the number
of spurious groups, we use the catalogue 
corresponding to a $\delta \rho/\rho$=200. For more details of the method, see \citet{Merchan:2005}. Those galaxies 
that are not in pairs or groups (number of members $\geq 3$) 
are also included in the M12 catalogue as single galaxies, i.e.
M12 lists all the DR12 galaxies in the redshift range $0.01\leq z \leq 0.3$. 
The number of single galaxies, pairs, and groups are 344732, 54373, and 31649, respectively. 

\subsection{Blazars}
We use the 5th edition of the Roma-BZCAT Catalog of Blazars (BZCAT, \citet{Massaro:2015}). The catalogue 
includes coordinates and multifrequency data for 3561 objects. The BZCAT includes a source classification and all 
have radio band detection, which means  that radio quiet BL Lac objects are not included. Blazers are classified as 
{\it BL Lac} and {\it BL Lac candidates}, {\it BL Lac-galaxy dominated} (with significant emission of the host galaxy over the nuclear 
emission), {\it QSO radio loud with Flat Spectrum}, and {\it blazars of uncertain classification}.  
Even though we are interested in BL Lac objects, we have considered all the sources in the BZCAT. Owing to the featureless 
optical spectrum,  many BL Lac objects have no redshift determination. Nevertheless, 748 (52\%) of the 1425 blazars classified as 
{\it BL Lac}, {\it BL Lac-candidate}, or {\it BL Lac-galaxy dominated} in the BZCAT have spectroscopic redshift determinations.

\begin{figure}
   \centering
   \includegraphics[width=9cm]{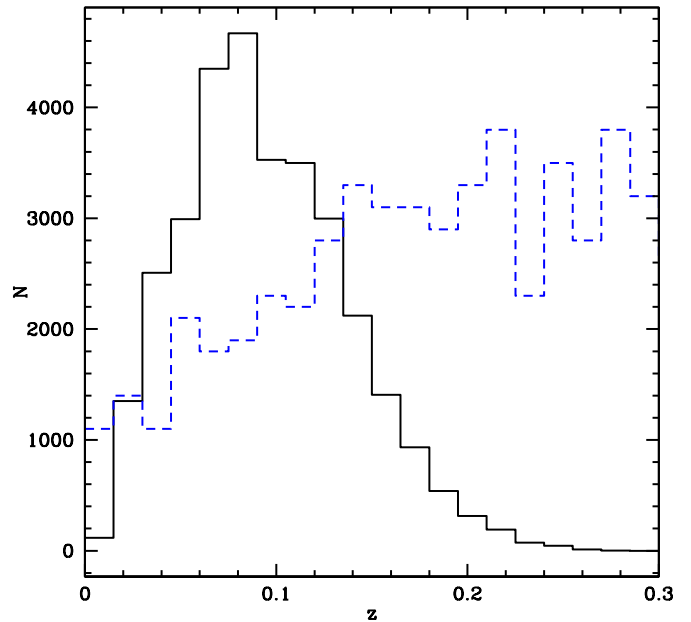}
   \caption{Redshift distribution of groups of galaxies (solid line) and for blazars in the BZCAT (blue dashed line). 
The numbers in the Y-axes correspond to the groups of galaxies; the number of blazars has been multiplied by 100 
to allow a better comparison.
   }
            \label{fig_z}%
    \end{figure}

Although the sample of blazars is deeper than the group catalogue, there are many blazars in the common redshift region. 
Figure \ref{fig_z} shows the redshift distribution of both blazars with redshift in the BZCAT and groups of galaxies with 
three or more members in the M12 catalogue.  

\section{Blazars in groups of galaxies}
\begin{figure}
   \centering
   \includegraphics[width=9.5cm]{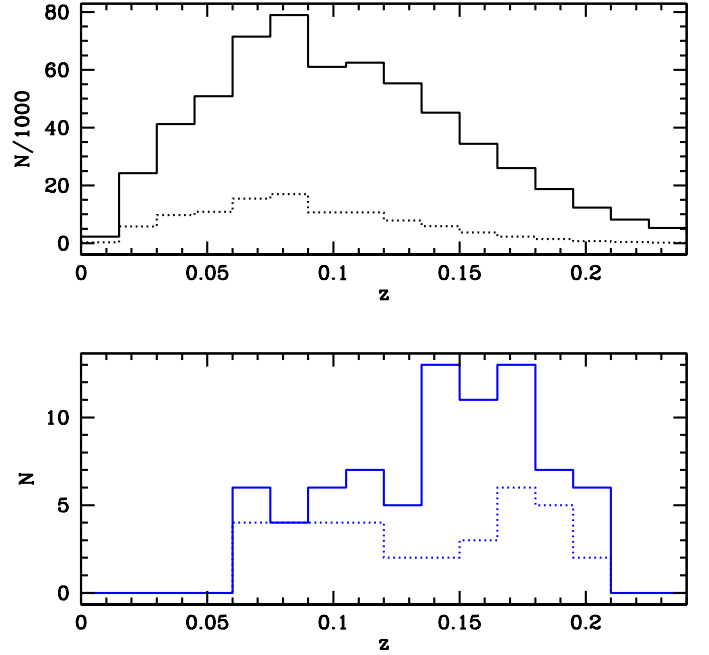}
   \caption{(upper panel) Redshift distribution of galaxies in the M12 catalogue (continuous line). The dotted 
line shows the redshift distribution of galaxies that are in groups of three or more members. (lower panel) As in the upper panel 
for BL Lac objects in the M12 catalogue.
   }
          \label{fig_z_II}%
    \end{figure}

Although the sample of blazars in the BZCAT is not complete in any sense, objects were not selected based on their 
environment. Consequently, the cross-correlation between blazars and groups of galaxies is free of selection bias.  
Several authors cross-correlated 
samples of blazars with optical sources in the SDSS (\citet{Collinge:2005}, \citet{Turriziani:2007}, \citet{Massaro:2014}).
In this work, we do not attempt to identify new BL Lac objects in the SDSS survey, nevertheless, since the BZCAT does not 
include the SDSS counterpart, a cross-identification is required. 
The position accuracy of the BZCAT is typically <1", but it could be of the order of 5" (\citet{Massaro:2014}. 
Using a linking length of 5", we cross-correlate all the galaxies and blazars in the 
M12 and the BZCAT, respectively. We found 121 galaxies in the M12 that have a counterpart in the BZCAT.  We compared 
the redshifts in both catalogues and confirmed that they correspond to the same objects. Fifty-five ($45 \pm 4\%$) 
of these galaxies appear as single galaxies in the M12 catalogue (no  companion galaxy according to the linking 
lengths of the FOF algorithm), 27 are in pairs, 
and 39 are in groups (three or more members). 
According to the type classification of \citet{Massaro:2014}, we found, as expected, that most 
identifications correspond to {\it BL Lac-galaxy dominated objects} (96);  the remaining blazars are {\it BL Lac} 
(24) and  {\it Blazar Uncertain type} (1). Consequently, hereafter we  refer to our sample as BL Lac objects. 
The lower panel of figure \ref{fig_z_II} shows the BL Lac objects with a counterpart in the M12 catalogue. The continuous 
line shows all the coincidences, and the dashed line corresponds to BL Lac objects in 
groups. As can be seen, the fraction of BL Lac objects in groups decreases with redshift and, moreover, there are no BL 
Lac objects in groups beyond $z\sim 0.23$.
This is due to the growth of incompleteness as a function of redshift that the FOF algorithm presents. This can be clearly 
seen in the  upper panel of figure \ref{fig_z_II}, where the same effect is present when galaxies in general are considered.
In a first attempt to deal with this incompleteness, we have restricted our analysis to BL Lac objects and galaxies with $z\leq 0.2$.
In this case, the total number of coincidences is 78, of which 25 correspond to single galaxies and 36 are in groups. 

In addition to the incompleteness discussed in the previous paragraph, there is another systematic effect that must be 
considered before computing the fraction of BL Lac objects that reside in groups. Owing to peculiar velocities, catalogues of groups of galaxies 
identified in redshift space tend to include a fraction of spurious groups. \citet{Merchan:2002} used a set of mock catalogues 
constructed from numerical simulations to estimate the fraction of spurious groups that arise when a FOF algorithm is applied.
They found that approximately 6\% of groups with four or more members and 13\% of triplets are spurious. In order to account 
for this effect, we implemented a set of Monte Carlo simulations to compute the fraction of BL Lac in groups of galaxies.
Using the bootstrap resampling technique we selected 1000 samples of groups excluding at random the expected fraction of spurious 
groups. With this procedure, we found that the percentage of BL Lac in groups of galaxies is $43 \pm 5\%$. We note that this value was 
obtained by disregarding the effect of incompleteness that also affects the identification of groups of galaxies in the redshift space.
 \citet{Merchan:2002} found that this incompleteness is at least 10\%. Moreover, the incompleteness grows with redshift. Evidence 
of this is the fact that, when we restrict our analysis to $z\leq 0.1$, the percentage of BL Lac objects in groups grows 
to $78 \pm 12 \%$. Figure \ref{fig_z}
clearly shows that the number density of groups of galaxies begins to fall at redshift $\sim0.07$ and few groups are 
identified at redshift > 0.2. Assuming that this incompleteness equally affects galaxies and BL Lac objects, we have selected 
a random sample of galaxies from the M12 catalogue that have the same redshift distribution of BL Lac objects with a counterpart in the 
M12. For this random sample, we found that the fraction of galaxies in groups 
is $19.3 \pm 0.1\%$, much smaller than the  $43\%$ found for BL Lac objects in groups. This result clearly shows that this type of AGN 
has a higher probability of being in groups than galaxies in general. 

The small number of BL Lac objects with a counterpart in the M12 catalogue does not permit the fraction 
of BL Lac objects in groups
as a function of redshift to be appropriately tested. Nevertheless, we can quantify the consequences of the incompleteness by using galaxies in the M12 
catalogue. The procedure is the 
following: we compute the fraction of galaxies in the M12 catalogue as a function of redshift in bins of $\Delta z=0.015$.
Table \ref{tab_zbin} shows that the percentage of galaxies in groups decreases with redshift from $\sim 30\%$ at low redshift 
($Per_{Low \hskip 0.05cm z}$) to $11.3\%$ for $z=0.2$. In this context, the percentage reported of 19.3\% ($Per_{0.0\le z\le 0.2 }$) of 
galaxies in groups in the redshift range 
$0\leq z \leq 0.2$ would be the average of these two values.
Assuming that 
the sample of groups in the M12 catalogue is complete at low redshift and that the fraction of galaxies that belong to groups does not depend
on redshift, we can correct the percentage of galaxies in groups up to z=0.2 as 
$$ CF=\frac { { Per }_{Low \hskip 0.05cm z } }{ { Per }_{ 0.0\le z\le 0.2 } } =1.55 \pm 0.01 $$ (the uncertainty was estimated by error 
propagation).
If we apply this correction factor to the observed fraction
of BL Lac objects in groups, 
we find that the expected percentage of BL Lac objects in groups 
is $\sim 67 \pm 8 \%$ (the uncertainty was estimated by error propagation). Considering that even the sample of groups at low redshift 
used to compute $CF$ may be incomplete, this  value should be taken as a lower limit.
We note that the  percentage is consistent with the $78 \pm 12\%$ found when restricting our analysis to
$z \leq 0.1$; however, this number was obtained with only 13 objects.

   \begin{table}
      \caption[]{Fraction of galaxies in groups as a function of redshift in the M12 catalogue.}
         \label{tab_zbin}
     $$ 
         \begin{array}{p{0.5\linewidth}r}
            \hline
            \noalign{\smallskip}
            Redshift      &  $\%$  \\
            \noalign{\smallskip}
            \hline
            \noalign{\smallskip}
        0.020 & 28.8 \pm 0.3 \\
        0.035 & 30.2 \pm 0.2 \\
        0.050 & 28.3 \pm 0.2 \\
        0.065 & 27.4 \pm 0.2 \\
        0.080 & 28.2 \pm 0.2 \\
        0.095 & 25.3 \pm 0.2 \\
        0.110 & 23.6 \pm 0.2 \\
        0.125 & 21.7 \pm 0.2 \\
        0.140 & 20.2 \pm 0.2 \\
        0.155 & 17.7 \pm 0.2 \\
        0.170 & 15.8 \pm 0.2 \\
        0.185 & 13.3 \pm 0.2 \\
        0.200 & 11.3 \pm 0.3 \\
            \noalign{\smallskip}
            \hline
         \end{array}
     $$ 
   \end{table}

\section{Conclusions}

Using a large sample of groups of galaxies, we estimated the percentage of BL Lac objects that are in groups. 
The sample of groups was selected by Merchan 2015 (M12), applying a {friends-of-friends}
algorithm in the SDSS-DR12. Those galaxies in the SDSS-DR12 that are not in pairs or groups are also included in the M12 sample 
as single galaxies. Galaxies in the M12 sample were cross-correlated with the sample of blazars in the BZCAT of \citet{Massaro:2015}.
We found that 121 galaxies in the M12 catalogue are blazars in the BZCAT sample. According to \citet{Massaro:2015}, 
these blazars have been classified as {\it BL Lac-galaxy dominated} (96), {\it BL Lac} (24), and  {\it Blazar Uncertain type} (1).  

Since the sample of groups of galaxies is highly incomplete beyond $z\sim 0.2$, we have restricted the analysis to this 
redshift. In this case, the number of coincidences between the two catalogues is 78. Once contamination by spurious groups 
is taken into account, $32 \pm 4 \%$ of the BL Lac objects
correspond to single galaxies and $43 \pm 5\%$ are in groups of three or more members. For a random sample of galaxies with the same 
redshift distribution, the percentage of galaxies in groups is $19.3 \pm 0.1\%$. Since samples of groups of galaxies selected using
FOF algorithms have a growing incompleteness with redshift, 
the number of BL Lac/galaxies in groups should be taken as lower limits. 
By analysing the fraction of galaxies in groups as a function of
redshift, we estimated a correction factor that takes into
account the incompleteness of the group catalogue. Once this
factor is applied to the sample of 78 BL Lac objects with counterpart in the M12 catalogue, 
we conclude that the percentage
of BL Lac objects in groups is $\gtrsim 67 \pm 8 \%$. Although this result
corresponds mainly to BL Lac-galaxy dominated objects and in
the nearby Universe, we see no reason why this result cannot be
extended to higher redshift or BL Lac objects in general. Interestingly, this fraction is similar to the value found for 
FR I galaxies in almost the same redshift range. \citet{Zirbel:1997} 
found that about 70\% of FR I galaxies at $z < 0.25$ reside in overdense environments while 
this is much less frequently the case for FR II galaxies that are believed to be the parent population of 
flat-spectrum radio loud quasars (see \citet{Castignani:2014} and \citet{Malavasi:2015} for high redshift objects).

The high rate of BL Lac objects in groups found in this work
strongly supports the method proposed by \citet{Muriel:2015} and
successfully used by \citet{Rovero:2016}.

\begin{acknowledgements}
We thank the anonymous referee for useful
questions and suggestions that improved this paper.
This work has been partially supported with grants from the
Consejo Nacional de Investigaciones Cient\'ificas y T\'ecnicas de la Rep\'ublica
Argentina (CONICET) and the Secretar\'ia de Ciencia y Tecnolog\'ia de la Universidad
de C\'ordoba. We thank M. Merch\'an for making publicly available the catalogue of 
groups of galaxies. Funding for the Sloan Digital Sky Survey (SDSS) was provided
by the Alfred P. Sloan Foundation, the Participating Institutions, the National
Aeronautics and Space Administration, the National Science Foundation, the
U.S. Department of Energy, the Japanese Monbukagakusho, and the Max
Planck Society. The SDSS Web site is http://www.sdss.org/. The SDSS is managed
by the Astrophysical Research Consortium (ARC) for the Participating
Institutions. The Participating Institutions are the University of Chicago,
Fermilab, the Institute for Advanced Study, the Japan Participation Group, Johns Hopkins University, the Korean Scientist Group, Los Alamos National
Laboratory, the Max Planck Institut f\"ur Astronomie (MPIA), the Max Planck
Institut f\"ur Astrophysik (MPA), New Mexico State University, the University of
Pittsburgh, the University of Portsmouth, Princeton University, the United States
Naval Observatory, and the University of Washington. 
\end{acknowledgements}

\bibliographystyle{aa} 
\bibliography{28736_hk} 

\end{document}